\documentclass[pop,fleqn]{w-art}
\usepackage{times}
\usepackage{w-thm}
\usepackage{graphicx}

\providecommand{\href}[2]{#2}

\usepackage{amsfonts}
\usepackage{amssymb}
\usepackage{cite}
\usepackage{epsfig}
\usepackage{psfig}

\newcommand{\ch}{{\rm ch}}
\newcommand{\be}{\begin{equation}}
\newcommand{\ee}{\end{equation}}
\newcommand{\bea}{\begin{eqnarray}}
\newcommand{\eea}{\end{eqnarray}}

\newcommand{\3}{{\bf 3}}
\newcommand{\2}{{\bf 2}}

\newcommand{\Anti}{{\bf Anti}}
\newcommand{\Sym}{{\bf Sym}}
\newcommand{\Adj}{{\bf Adj}}
\newcommand{\n}{{\bf n}}
\newcommand{\N}{{\bf N}}

\newcommand{\ov}{\overline}
\def\IR{\relax{\rm I\kern-.18em R}}

\def\IP{\relax{\rm I\kern-.18em P}}
\def\inbar{\vrule height1.5ex width.4pt depth0pt}
\def\IC{\relax\,\hbox{$\inbar\kern-.3em{\rm C}$}}

\def\K3{{\bf K3}}

\def\a{\alpha}
\def\b{\beta}
\def\d{\delta}
\def\g{\gamma}

\def\ov{\overline}

\begin{document}
\pagespan{1}{}



\begin{flushright} \vspace{-2cm}
{\small MPP-2005-163\\
LMU-ASC 79/05} 
\end{flushright}
\title[Heterotic Vacua from general (non-) Abelian Bundles]{Heterotic Vacua from general (non-) Abelian Bundles }


\author[T.Weigand]{Timo Weigand\inst{1,2}%
\footnote{\textsl{weigand@mppmu.mpg.de}}}
\address[\inst{1}]{Max-Planck-Institut f\"ur Physik\\F\"ohringer Ring 6, 80805 M\"unchen, Germany}
\address[\inst{2}]{Arnold-Sommerfeld-Center for Theoretical Physics\\Department f\"ur Physik, Ludwig-Maximilians-Universit\"at  M\"unchen\\Theresienstra{\ss}e 37, 80333 M\"unchen, Germany}

\begin{abstract}
We report on the construction of four-dimensional string vacua by considering general abelian and non-abelian bundles on an internal Calabi-Yau for both heterotic theories. The structure of the resulting gauge sector is extremely rich and gives rise to many new model building possibilities. We analyse the chiral spectrum including the contribution from heterotic five-branes and provide the general consistency conditions. The one-loop corrected supersymmetry condition on the bundles is found to be that of $\pi$-stability. As an application we present a supersymmetric Standard-Model like example for the $SO(32)$ string with $U(n)$ bundles on an elliptically fibered Calabi-Yau.  
\end{abstract}
\maketitle 




\renewcommand{\leftmark}
{T. Weigand: Heterotic Vacua from general (non-) Abelian Bundles}

\noindent Talk given at the RTN network conference {\emph {Constituents, Fundamental Forces and Symmetries of the Universe} in Corfu, Greece, 20-26 September 2005.
To appear in Fort. Phys.

\section{Introduction and summary}

Even if we are still far from a complete understanding of all the mysteries string theory confronts us with, it is widely accepted by now that the theory gives rise to an overwhelmingly large number of at least meta-stable vacua. Each of them can be approximately described by a low-energy effective theory which, unlike adhoc phenomenologically motivated guesses of effective lagrangians, is guaranteed to possess a consistent ultra-violet completion. A thorough analysis of the set of possible string vacua is therefore an equally challenging and indispensible endeavour. The present state of the art only admits the description of a tiny fraction of possible constructions, the mathematics of the emergent structure being simply too intricate in most cases. One large class of solvable string compactifications with phenomenologically appealing features is given by type IIA toroidal orientifolds with intersecting $D6$-branes (for the most recent review see \cite{Blumenhagen:2005mu}). In view of the intense model building attempts in this framework, it would be very desirable to extend this type of models to the generic case of curved internal background manifolds. Progress in this direction is severely hampered by our lacking knowledge of general supersymmetric cycles on which the branes wrap since the special Lagrangian condition, being non-holomorphic in nature, is of course beyond the power of complex geometry. On the other hand, mirror symmetry maps intersecting branes to B-type branes on holomorphic cycles carrying {\emph{abelian}} gauge instantons. The sLag condition gets replaced by a stability constraint on the holomorphic gauge bundle. An immediate two-fold extension of the intersecting branes picture is therefore to consider {\emph{general}} $U(n)$ bundles on $D9$-branes wrapping a {\emph{non-trivial}} Calabi-Yau. This setup is in turn reminiscent of the S-dual heterotic string constructions. Here, by contrast, most attempts in the literature have focused on the embedding of non-abelian vector bundles with vanishing first Chern classes into $E_8 \times E_8$\footnote{For some early examples with $U(n)$ bundles see \cite{Green:1984bx,Strominger:1986uh,Andreas:2004ja,Aldazabal:1996du} and further references in \cite{Blumenhagen:2005zg}.}. 

In this talk we would like to report on recent activities in the exploration of four-dimensional vacua in both heterotic theories defined by allowing also for line and $U(n)$ bundles. In the $E_8 \times E_8$ case the use of line bundles yields a large number of new scenarios of breaking $E_8$ down to GUT groups or further to the Standard Model group {\emph {without the need of non-trivial Wilson lines on the Calabi-Yau}\cite{Blumenhagen:2005ga}. For the $SO(32)$ string {\cite{Blumenhagen:2005pm,Blumenhagen:2005zg} we will systematically give a class of $U(n)$ embeddings into $SO(32)$ which is the direct S-dual of magnetized $D9$-branes with non-abelian bundles \cite{Blumenhagen:2005zh}. We present the resulting chiral spectrum, taking also into account the possibility of allowing for additional heterotic five-branes. The absence of global and local anomalies imposes a number of non-trivial consistency conditions on the topological data. Besides being relevant for determinig the unbroken abelian part of the four-dimensional gauge group, the one-loop GS counterterms allow for a derivation of the corrections to the supersymmetry condition on the bundles at one loop. These arise as a D-term in the supergravity description. The result is that the perturbatively exact supersymmetry constraint is that of $\pi$-stability \cite{Enger:2003ue}, which, in the S-dual picture, constitutes the perturbative part of the well-known $\Pi$-stability condition. As an illustration of the rich model building possibilities behind these constructions we finally give an example in the $SO(32)$ framework by means of the spectral cover construction of stable vector bundles on an elliptically fibered Calabi-Yau. Many more of such phenomenologically interesting models are to be discovered. In particular, vacuum search in the direct Standard Model group breaking scenarios of  \cite{Blumenhagen:2005ga} on manifolds with trivial fundamental group might turn out fruitful.
Lack of space regrettably forces us to refer the reader to the original papers \cite{Blumenhagen:2005ga,Blumenhagen:2005pm,Blumenhagen:2005zg,Blumenhagen:2005zh}  for more detailed information and for a more complete list of relevant references.

\section {General embeddings and their chiral spectrum}

Four-dimensional heterotic compactifications are, at the perturbative level, specified by a vector bundle $W$
 over the internal Calabi-Yau manifold $\cal M$ together with an embedding of its structure group $G$ into 
$E_8 \times E_8$ or $Spin(32)/{\mathbb Z}_2$\footnote{In standard abuse of notation we will refer to this case as $SO(32)$ though the ${\mathbb Z}_2$ projection is actually a different one.}, respectively. 
By standard arguments, giving VEVs to the field strengths of $G$ on the internal manifold  breaks the
 four-dimensional gauge group to the commutator of  $G$ in $E_8 \times E_8$ or  $SO(32)$. 
In general we will be interested in Whitney sums of bundles $W = \bigoplus_{s} V_{s}$ with 
structure group $\prod_{s} G_s$.

Beyond the hitherto mostly studied case where all summand bundles are semi-simple, the 
$E_8 \times E_8$ string naturally allows for the embedding of  sums of $SU(n_i)$ vector bundles 
together with non-trivial line bundles, i.e. $W=\bigoplus_{i=1}^K  V_{i} \oplus \bigoplus_{m=1}^M  
L_m$, or modifications thereof\footnote{One can also take the $V_{i}$ to be $U(n_i)$-bundles and adjust the $c_1(L_m)$ to yield $c_1(W)=0$.}, as detailed in \cite{Blumenhagen:2005ga}. The concrete form of the 
four-dimensional gauge group depends very much on the details of the chosen embedding.
 In particular, concerning its abelian part, only the massless combinations of $U(1)$ 
factors survive as a gauge symmetry. 

For the $SO(32)$ theory a large class of models is defined by taking the structure group of $V_i$ to be
 $U(n_i)$ and diagonally embedding it into a $U(M_i)$ gauge factor with $M_i=n_i\, N_i, i \in \{1,\ldots,K\}$\cite{Blumenhagen:2005zg}. 
This breaks the original $SO(32)$ gauge group to $SO(2M) \times \prod_{i=1}^{K} U(N_i)$
with \mbox{$M+\sum_{i=1}^{K} N_i=16$}, again up to the issue of some subgroup of $U(1)^K$ potentially 
becoming massive. The subsequent notation will be referring to these two types of constructions in both 
heterotic theories.

To determine the massless spectrum, one analyses the splitting of the  adjoint respresentation of  
$E_8 \times E_8$ or $SO(32)$ into various irreducible representations $(R_j;r_j)$ under the four-dimensional 
group and the internal one. Thanks to the non-trivial internal gauge background we find four-dimensional 
{\emph {chiral}} matter in representations $R_j$ specified by the cohomology class $H^*({\cal M},U_j)$, where, loosely speaking, the 
bundle $U_j$ is determined as the one with fundamental representation $r_j$. In particular, the 
Hirzebruch-Riemann-Roch index theorem computes the net number of four-dimensional chiral fermions in the
representation $R_j$ as
\bea 
\label{RRH}
         \chi({\cal M},U_j)=\sum_{n=0}^3  (-1)^n \, {\rm dim}\, H^n({\cal M},U_j)
         =\int_{\cal M}\left[ {\rm ch}_3(U_j)+
           {1\over 12}\, c_2(T)\, c_1(U_j) \right].
\eea
Concretely, for the class of $SO(32)$ models we arrive at the perturbative massless spectrum
\bea
\label{SO32spec}
{\bf 496} \to
\left(\begin{array}{c} 
(\Anti_{SO(2M)})_0\\
\sum_{j=1}^{K} (\Adj_{U(N_j)};\Adj_{U(n_j)})\\
\sum_{j=1}^{K} (\Anti_{U(N_j)};\Sym_{U(n_j)}) + 
               (\Sym_{U(N_j)};\Anti_{U(n_j)}) + h.c.\\
\sum_{i < j} (\N_i,\N_j;\n_i,\n_j) + (\N_i,\ov{\N}_j,\n_i,\ov{\n}_j) + h.c. \\
\sum_{j=1}^{K} (2M, \N_j;\n_j) + h.c.\\
\end{array}\right).   
\eea
For examples in the context of the $E_8 \times E_8$ string we refer the reader to\cite{Blumenhagen:2005ga}.
   
In addition to this perturbative sector, both heterotic theories are well-known to comprise five-branes H5. 
For four-dimensional Lorentz invariance, they are taken to be spacetime-filling and wrap internal 2-cycles, 
which in supersymmetric configurations have to be holomorphic. In the $E_8$  theory, the worldvolume of the 
five-brane contains a tensor field and does not yield any additional charged chiral matter in four 
dimensions~\cite{Lukas:1998hk} in agreement with the observation that in Horava-Witten theory, the corresponding 
M5-brane can be pulled into the 11D bulk. We will therefore 
focus in the subsequent discussion on the $SO(32)$ case.
Here the worldvolume of a H5-brane supports a massless 
gauge field, which compared to the $E_8\times E_8$ H5-brane leads to  
different low energy physics. The gauge group can be deduced by noting that  S-duality directly maps the 
H5 to the D5-brane in Type I~\cite{Witten:1995gx}, which is known to give rise to symplectic gauge groups. More precisely, a 
brane wrapping  the holomorphic curve $\Gamma  = \sum_a N_a \,\Gamma_a , \, N_a \in {\mathbb Z}_0^{+}$ 
yields a gauge group factor $\prod_a Sp(2 N_a)$, where the enhancement is due to the multiple 
wrapping around each irreducible curve $\Gamma_a$. The above decomposition of $\Gamma$  may not be unique 
and the gauge group may therefore vary in the different regions of the associated moduli space. However, 
its total rank and the total number of chiral degrees of freedom charged under the symplectic groups are 
only dependent on $\Gamma$, of course.

The cancellation of gravitational anomalies on the $SO(32)$ H5-brane requires a Chern-Simons like 
coupling of the H5-brane to the bulk by anomalous inflow arguments. Inspired further by heterotic-Type I 
duality, one can infer that  the effective low energy action on the H5-branes
contains a piece
\bea
\label{CS}
S_{H5_a}= - \mu_5 \int_{\mathbb R_{1,3}\times \Gamma_a} \sum_{n=0}^1
B^{4n+2}\wedge \left( N_a+ {\ell_s^4\over 4(2\pi)^2} {\rm tr}_{Sp(2N_a)} F_a^2
\right) \wedge 
        { \sqrt{\hat{\cal A}({\rm T}\Gamma_a)}\over \sqrt{\hat{\cal A}({\rm N}\Gamma_a)} },
\eea
with the H5-brane tension  $\mu_5= \frac{1}{(2 \pi)^5 \,(\alpha')^3}$.
T$\Gamma_a$ and  N$\Gamma_a$ denote the tangent bundle and  the normal
bundle,  respectively, of the 2-cycle $\Gamma_a$, which  for concreteness we take to be irreducible from 
now on and wrapped by a stack of $N_a$ H5-branes. 
The curvature occurring in the definition of the $\hat{\cal A}$ genus is
defined as ${\cal R} = -i\ell_s^2 R \, (\ell_s \equiv 2 \pi \sqrt{\alpha'})$. 
Note that the universal five-brane 
coupling to $B^{(6)}$ (defined by $\ast_{10} dB^{(2)} = e^{2 \phi_{10}} d B^{(6)}$) 
must also be present for the $E_8 \times E_8$ theory irrespective of the concrete form of the further terms in its 
worldvolume action.  

To make contact with the vector bundle theoretic discussion of the massless spectrum, it is useful to 
describe the $SO(32)$ H5-brane wrapping $\Gamma_a$ as 
the skyscraper sheaf ${\cal O}\vert_{\Gamma_a}$ with toplogical invariants given by
 $\ch({\cal O}\vert_{\Gamma_a})=(0,0,-\gamma_a,0)$.
 
As anticipated, the $SO(32)$ H5-branes give rise to chiral
matter in the bifundamental $(\N_i, 2\N_a)_{1(i)}$, which is counted by the index
\bea
\label{ind}
\chi(X, V_i{\otimes \cal O}\vert^*_{\Gamma_a} )  = -\int_X c_1(V_i) \wedge \gamma_a.
\eea
Here $\gamma_a$ denotes the Poincar\'e dual 4-form corresponding to the 2-cycle $\Gamma_a$.

To conclude this summary of the heterotic particle content, Table~\ref{Tchiral1} exemplifies the chiral matter arising 
from the perturbative and non-perturbative sector of the $SO(32)$ theory with our favourite 
embedding (\ref{SO32spec}).

\begin{table}[htb]
\renewcommand{\arraystretch}{1.5}
\begin{center}
\begin{tabular}{|c||c|}
\hline
\hline
reps. & $H=\prod_{i=1}^K SU(N_i)\times U(1)_i \times SO(2M)\times \prod_{a=1}^L Sp(2N_a)$   \\
\hline \hline
$(\Adj_{U(N_i)})_{0(i)}$ & $H^*(X,V_i \otimes V_i^{\ast})$  \\
\hline
$(\Sym_{U(N_i)})_{2(i)}$ & $H^*(X,\bigwedge^2 V_i)$  \\
$(\Anti_{U(N_i)})_{2(i)}$ & $H^*(X, \bigotimes^2_s  V_i)$  \\
\hline
$(\N_i,\N_j)_{1(i),1(j)}$ & $H^*(X, V_i \otimes V_j)$ \\
$(\N_i,\ov \N_j)_{1(i),-1(j)} $ &  $H^*(X, V_i \otimes V_j^{\ast})$ \\
\hline
$(\Adj_{SO(2M)})$ & $H^*(X,{\cal O})$ \\
$(2{\bf M}, \N_i)_{1(i)}$ & $H^*(X, V_i)$\\
\hline
$(\Anti_{Sp(2N_a)})$ &  Ext$_X^*({\cal O}\vert_{\Gamma_a},{\cal O}\vert_{\Gamma_a})$ \\ 
$(\N_i,2\N_a)_{1(i)}$ & Ext$_X^*(V_i,{\cal O}\vert_{\Gamma_a})$  \\
$(2\N_a,2\N_b)$ & Ext$_X^*({\cal O}\vert_{\Gamma_a},{\cal O}\vert_{\Gamma_b})$  \\
\hline
\end{tabular}
\caption{\small Massless spectrum of the SO(32) theory with 
the structure group  taken to be 
$G=\prod_{i=1}^{K} U(n_i)$.  }
\label{Tchiral1}
\end{center}
\end{table}

\section{Anomalies, tadpoles and massive $U(1)$ factors}

In order to describe a well-defined string compactification the bundle data has to satisfy a number of 
non-trivial consistency conditions. A powerful guideline in their identification is 
to search for the appearance of possible global or local anomalies on the 
worldsheet or in the effective supergravity.
 
To begin with, the absence of anomalies in  the two-dimensional non-linear
sigma model~\cite{Witten:1985mj} requires
\begin{equation}
\label{K-theory}
c_1(W) \in H^2({\cal M}, 2{\mathbb Z}).
\end{equation}
In the $SO(32)$ theory, this constraint ensures that the number of chiral fermions in the fundamental of the 
$Sp(2 N_a)$ groups be even, as is obvious from (\ref{ind}), and therefore has a simple interpretation as 
the condition for the vanishing of a global $Sp(2 N_a)$ Witten anomaly on every probe brane.

Let us turn our attention to the local anomalies of the spacetime effective theory.
It is well-known that for absence of anomalies in ten dimensions the string tree-level effective action has to be modified by two important contributions:
First, the three-form field strength comprises not only gauge, but also essential gravitational CS terms, 
$H=dB^{(2)}-{\alpha'\over 4}(\omega_{Y}-\omega_{L})$, which enter into the effective action 
via the usual kinetic term $S_{kin}=-{1\over 4\, \kappa_{10}^2}\, \int  e^{-2\phi_{10}}\,
       H\wedge \star_{10}\, H$
with $\kappa^2_{10}={1\over 2}(2\pi)^7\, (\alpha')^4$.
We immediately see that the Bianchi identity for  $H_3$ takes the form
$dH_3={\ell_s^2}\left( {1\over4\,  (2\pi)^2}\left[ 
{\rm tr} R^2 -{\rm tr} F^2 \right]+ \sum_a N_a \,\gamma_a   \right)$ 
after including also the CS coupling \ref{CS} of the dual field 
$B^{(6)}$ to the H5 brane.
The traces are taken in the fundamental representation of $SO(1,9)$ and $E_8 \times E_8$ or $SO(32)$, 
respectively.
This translates into the cohomological condition
\bea 
\label{TP}
-c_2(T)+ \sum_{a=1}^L N_a \gamma_a= 
\left\{\begin{array}{lc}
\sum_{i=1}^K {\rm ch}_2(V_{n_i}) + \sum_{m=1}^M  a_m\, c^2_1(L_m) & (E_8 \times E_8), \\ 
\sum_{i=1}^{K} N_i \; \ch_2(V_i)  &  (SO(32)). 
\end{array} \right. 
\eea
Note that in the first case the coefficients $a_m$ depend on the concrete embedding and 
are determined as $4a_m=\rm{tr}_{E_8}Q^2_m$.
Besides its interpretation as a tadpole condition in supergravity, 
this constraint ensures, in the $SO(32)$ case, the absence of 
cubic non-abelian anomalies\footnote{Note, however, that in 
the $E_8 \times E_8$ embeddings of \cite{Blumenhagen:2005ga} these non-abelian anomalies
vanish even without imposing (\ref{TP}). By contrast, in both heterotic theories it is 
required for the generalized 
Green-Schwarz mechansim to cancel all (possibly mixed) abelian anomalies}.

Since it will become relevant in the next section, we would like to draw the reader's attention to the fact that the crossterm
$S_{kin} = \frac{1}{8\pi \ell_s^6} \int \left( {\rm tr} F^2 - {\rm tr} R^2 \right)\wedge B^{(6)}$  
contained in the kinetic action of $H$ effectively appears at one loop in string perturbation theory.

The second one-loop piece of information provided by anomaly counterterms is of course the 
celebrated ten-dimensional
Green-Schwarz term 
$S_{GS} = \frac{1}{48(2\pi)^3 \ell_s^2} \int B^{(2)} \wedge X_8$
with the standard anomaly eight-form. 
As shown in great detail in \cite{Blumenhagen:2005ga,Blumenhagen:2005pm} dimensional reduction of the kinetic and GS-terms 
provides precisely the right counterterms to cancel all abelian anomalies arising 
in four dimensions.     
Rather than displaying all terms here, we would like to stress that one important ingredient 
in these four-dimensional 
counterterms are linear couplings of the abelian field strength to the various two-forms 
arising from reduction of $B^{(2)}$ and $B^{(6)}$ on a basis of internal two-cycles 
$\omega_k$ and their dual four-cycles $\hat{\omega}_k$ as  
$B^{(2)} = b^{(2)}_0+\ell_s^2\, 
          \sum_{k=1}^{h_{11}}   b^{(0)}_k\, \omega_k ,\quad  B^{(6)}=\ell_s^6\, b^{(0)}_0\,  {\rm vol}_6 +
      \ell_s^4\,
           \sum_{k=1}^{h_{11}}  b^{(2)}_k\, \widehat\omega_k$.
Collecting all contributions of this type 
one finds
\begin{equation}
\label{massterms}                 S_{mass}=\sum_{x}  \sum_{k=0}^{h_{11}} 
            {Q^x_k\over 2\pi\alpha'}
                \int_{\IR_{1,3}} f_x \wedge b^{(2)}_k,
\end{equation}
where the abelian field strengths are collectively denoted by $f_x$ and for brevity we refer again 
to \cite{Blumenhagen:2005ga,Blumenhagen:2005pm} 
for the concrete formluae for $Q$ in both heterotic theories.
The couplings (\ref{massterms}) induce a mass for every linear combination of $U(1)$s which does not lie
in the 
kernel of the mass matrix $Q$.
Let us point out that all mass terms are of the same order in both string
and sigma model perturbation theory. The number of massive $U(1)$s will always be at least as big as
the number of anomalous $U(1)$s.
Though all entries in the mass matrix are of order $M_s^2$, 
the mass eigenstates  of the gauge bosons can have
masses significantly lower than the string scale.

\section{$\mathbb{\pi}$-stability for supersymmetry at one loop}
Supersymmetry at string tree-level imposes the well-known constraint $g^{ab} F_{ab}=0$ on the field strength 
$F$ of $W$,  or equivalently that F be holomorphic, $F^{(2,0)}=0=F^{(0,2)}$, and satisfy 
$J \wedge J \wedge F=0$. 
The Donaldson-Uhlenbeck-Yau theorem translates this latter condition, 
the zero-slope limit of the Hermitian Yang-Mills equation, 
into the requirement that each summand bundle $V_i$ be $\mu$-stable and satisfy the integrability condition 
$\int_{\cal M} J \wedge J \wedge c_1(V_i)=0$, which is of course trivially fulfilled for gauge bundles with vanishing first Chern class.
Being a non-holomorphic supersymmetry constraint, 
this so-called DUY condition is naturally expected to arise as a D-term in the four-dimensional effective 
supergravity and is therefore, at the string perturbative level, subject to at most one-loop corrections. 
Fortunately, we actually have just encountered important one-loop terms 
in our effective action by standard anomaly considerations, the mass couplings
analysed in the previous section. They carry the relevant information to 
determine the potential one-loop corrections to the DUY equation. The key point is to notice that the 
linear mass couplings involve the axionic partners complexifying the dilaton and K\"ahler moduli as
$S={1\over 2\pi}\left[ e^{-2\phi_{10}} {{\rm Vol}({\cal M}) \over
             \ell_s^6 } + {i}\, b^{(0)}_0 \right]$ and $T_k={1\over 2\pi}\left[ -\alpha_k + {i} b^{(0)}_k \right]$.
To maintain gauge invariance in the K\"ahler potential $\cal K$ of the effective ${\cal N}=1$ supergravity,
 the couplings (\ref{massterms}) also enforce
an appropriate modification of $\cal K$ as 
\bea
{\cal K}&=&{M^2_{pl}\over 8\pi} \Biggl[   
       -\ln\Biggl(S+S^*-\sum_x Q^x_0\, V_x\Biggr)-\ln\Biggl(-\sum_{i,j,k=1}^{h_{11}}
{d_{ijk}\over 6} 
\biggl(  T_i+T_i^*-\sum_x Q^x_i\, V_x\biggr) \nonumber \\
&& \phantom{aaaaaaa} \biggl(  T_j+T_j^*-\sum_x Q^x_j\, V_x\biggr)
\biggl(  T_k+T_k^*-\sum_x Q^x_k\, V_x\biggr) \Biggr) \Biggr],
\eea
where by $V_x$ we denote the abelian superfields.
As a standard matter of fact, the ${\cal N}=1$ K\"ahler potential is related to the Fayet-Iliopoulos D-terms 
terms via the relation      
${\xi_x\over g_x^2}=   {\partial {\cal K}\over \partial V_x} \biggr\vert_{V=0}$.
We therefore arrive at the following tree-level and one-loop contributions to the FI terms
\bea
 {\xi_x\over g_x^2} = -\frac{e^{2 \phi_{10}} M_{pl}^2 \,\ell_s^6}{4  {\rm Vol}({\cal M})}
\Bigl(\frac{1}{4} e^{-2 \phi_{10}} \sum_{i,j,k=1}^{h_{11}} d_{ijk} Q^x_i \alpha_j \alpha_k - \frac{1}{2} Q^x_0 \Bigr).
\eea
Inserting the concrete expressions for the charges we find that the FI terms vanish
if and only if\footnote{We assume here that the VEVs of the matter fields charged under the abelian gauge group are zero.}  
\begin{equation}
\label{DUYloopE}
\begin{array}{l}
\int_{\cal M} J\wedge J \wedge c_1(L_n)  -  \frac{1}{2}\,g_s^2\, \ell_s^4 \,
      \int_{\cal M} c_1(L_n) \wedge \left(
      \sum_{i=1}^K {\rm ch}_2(V_{i}) + 
        \sum_{m=1}^M  a_m\, c^2_1(L_m) +{1\over 2}\, c_2(T)\right)=0, \nonumber\\
\int_{\cal M} J\wedge J \wedge c_1(V_i)  - 
  2\, g_s^2\, \ell_s^4 \, \int_{\cal M} 
\left( {\rm ch}_3 (V_i)+\frac{1}{24}\, c_1(V_i)\,  c_2(T)\right)=0 \quad\quad\quad\quad\quad\quad\quad\quad\quad\quad\quad (10)
\end{array}
\end{equation}
for the $E_8 \times E_8$ and $SO(32)$ case, respectively.  
Note that in the first case, the one-loop term contains a sum over all bundles in the same $E_8$ factor as $L_n$, whereas for the $SO(32)$ string it is "local" in that it only depends on the bundle $V_i$ under consideration. 
Since the tree-level part of this expression constitutes the conventional DUY equation, 
we interpret (10) as the one-loop corrected integrability condition for the local 
supersymmetry equation. Arising as a D-term, it poses constraints on a particular combination of the dilaton and the K\"ahler moduli. For consistency these values have to lie in the perturbative regime and inside the K\"ahler cone.
  As recalled above, to be also {\emph{sufficient} for supersymmetry, the DUY equation has to be supplemented
by an appropriate stability condition. In fact, we argued in \cite{Blumenhagen:2005pm} that
the modified stability condition
to be satisfied by each subbundle in addition to (10) is precisely that of  $\pi$-stability 
\cite{Enger:2003ue}. For practical applications it is satisfactory to note that,
 at least in the perturbative regime, 
$\mu$-stable bundles are also $\pi$-stable \cite{Enger:2003ue}, but since the converse is not true 
it would be interesting to investigate the moduli space of bundles acceptable only under the 
latter notion of stability.
 Let us simply state here that a further relevant effect of the anomaly counterterms is 
to generate non-universal one-loop corrections to the 
gauge kinetic functions~\cite{Blumenhagen:2005ga,Blumenhagen:2005pm}. For consistency of the supersymmetric
framework we have to ensure that their real part is still positive, i.e.
\begin{equation}
\label{DUYloopE}
\displaystyle
\begin{array}{l}
\int_{\cal M} J\wedge J \wedge J -  \frac{3}{2}\,g_s^2\, \ell_s^4 \,
      \int_{\cal M}  \,J \wedge  \left(\frac{4}{3}\, a_n \, c_1^2(L_n) + 
      \sum_{i=1}^K {\rm ch}_2(V_{i}) + 
        \sum_{m=1}^M  a_m\, c^2_1(L_m) +{1\over 2}\, c_2(T)\right) > 0, \nonumber\\
{n\over 3!}\,
 \int_{\cal M} J \wedge J \wedge J -
g_s^2\, \ell_s^4\,  \int_{\cal M} J \wedge 
\left( {\rm ch}_2 (V_i)+\frac{n}{24}\, c_2(T)\right) >0, \quad\quad\quad\quad\quad\quad\quad\quad\quad\quad\quad (11)
\end{array}
\end{equation} 
respectively\footnote{The first inequality has to be satisfied by each line bundle $L_n$ in the $E_8 \times E_8$ models; unlike with (10), the analogous condition on the $V_i$ is non-trivial and simply given by omitting the term $\frac{4}{3} \, a_n c_1^2(L_n)$.}. One can then verify that S-duality translates the complete supersymmetry 
constraint of the $SO(32)$ theory precisely into the perturbative part of the celebrated $\Pi$-stability 
condition for spacetime-filling D-branes. Of course, (10) and (11) may receive non-perturbative corrections in $\alpha'$ and $g_s$. In view of S-duality, it is therefore desirable to 
investigate the possible analogue of full  $\Pi$-stability for both heterotic theories.

\section{A Standard-Model like example on an elliptically fibered Calabi-Yau}

As a demonstration of the new model building possibilities let us present a simple example with Standard Model-like gauge group and chiral matter in the $SO(32)$ string.  
As has become clear, the general strategy is to construct \emph{stable holomorphic} vector 
bundles with known topological invariants on 
a given Calabi-Yau and to ensure that the various 
consistency conditions (\ref{K-theory},\ref{TP},10,11) are satisfied. It is then possible to engineer
systematically interesting semi-realistic low-energy properties. Fortunately, 
the so-called spectral cover construction provides us with a large class of such stable $SU(n)$ bundles
 on elliptically
fibered Calabi-Yaus. These bundles can then be further twisted by line bundles to yield 
structure group $U(n)$ as needed for our $SO(32)$ models. We refer the reader to
 \cite{Friedman:1997yq} for the details of the spectral cover construction; here we can merely recall the 
main ingredients. 
Consider an elliptically fibered Calabi-Yau threefold ${\cal M}$ with projection $\pi:{\cal M}\to B$
 and a section $\sigma:B\to {\cal M}$ which identifies the base $B$ as a submanifold of ${\cal M}$.
If the base is smooth and preserves only ${\cal N}=1$ supersymmetry
in four dimensions, it is restricted to a del Pezzo surface, a Hirzebruch
surface, an Enriques surface or a blow up of a Hirzebruch surface.
The idea is to use a simple description of $SU(n)$ bundles over
the elliptic fibers $E_b$ over each base point $b$ and then globally glue them together to define
bundles over ${\cal M}$. One of the many nice properties of such a construction is
that eventually the Chern classes of such bundles can be computed 
entirely in terms of data defined on the base $B$. The $SU(n)$ bundle is specified partly by the spectral cover
 $C$, an $n$-fold cover of $B$ with $\pi_C:C\to B$. One has the freedom to further twist it by $\eta$,  
the pull-back of a line bundle  on the base. This determines the cohomology class of $C$ 
as $[C]=n\sigma + \eta$. In addition 
one has to choose a line bundle ${\cal N}$ on $C$ defined such that $V\vert_B=\pi{_C*}{\cal N}$.
The vector bundle $V$ is then given as $V=\pi_{1*}(\pi_2^*\, {\cal N}\otimes {\cal P})$,
where $\pi_1$ and $\pi_2$ denote the two projections of the fiber product $Y={\cal M}\times_B C$ onto
the two factors ${\cal M}$ and $C$ and ${\cal P}$ represents the Poincar\'{e} bundle on $Y$. 
If the spectral cover is irreducible, i.e. if the linear system $|\eta|$ is 
basepoint-free and $\eta-n\, c_1(B)$ is effective \cite{Donagi:2004ia}, then the in general semi-$\mu$-stable bundle 
$V$ is truly stable.   
To arrive at a $U(n)$ bundle, we twist $V$ by an arbitrary line bundle ${\cal Q}$ on ${\cal M}$ with 
$c_1({\cal Q})=q\, \sigma + c_1(\zeta)$ to get $ V_{{\cal Q}}=V\otimes {\cal Q}$. Note that the process of 
twisting does not affect the stability properties of the bundle. For the very explicit expressions of the 
Chern classes of $V_{{\cal Q}}$ in terms of the above data please consult \cite{Blumenhagen:2005zg}. 
Suffice it here to state that they are determined entirely by the rank $n$ of $V$, $c_1({\cal Q})$, $\eta$ and a
further rational number $\lambda$ which has to be chosen appropriately to guarantee integer expressions
for the Chern classes.

Let us now outline a concrete example in the framework of the $SO(32)$ theory
 on an elliptic fibration over the del Pezzo surface $B={\rm}dP_4$. More details on the computations can be found in \cite{Blumenhagen:2005zg}. 
The second cohomology class of ${\rm}dP_4$
is generated by the elements $l,E_1,\ldots,E_4$ with intersection form
$l\cdot l=1, \quad l\cdot E_m=0, \quad E_m\cdot E_n=-\delta_{m,n}$.
Being interested in Standard-like models we  aim at obtaining  a visible gauge group 
$U(3)_{\a}\times U(2)_{\b} \times U(1)_{\g} \times U(1)_{\d}$ and at realizing 
the quarks and leptons as appropriate bifundamentals. 
A possible  choice of the hypercharge as a (massless) 
combination of the abelian factors is  given by  $Q_Y= \frac{1}{6} Q_{\a} + \frac{1}{2}(Q_{\g} + Q_{\d})$. 
In this case, also some of the  (anti-)symmetric representations carry MSSM quantum numbers. 
The details of the chiral MSSM spectrum we try to reproduce can be found in Table~\ref{MSSM}.
Among the many possibilities we consider the simple embedding of the structure group 
$G=U(1)_a \times U(1)_b \times U(2)_c \times U(1)_d$ into $U(3) \times U(2) \times U(2) \times U(1)$. This breaks $SO(32)$ to the commutator 
$U(3)_{\a}\times U(2)_{\b} \times U(1)_{\g} \times U(1)_{\d} \times SO(16)$
modulo the issue of anomalous abelian factors. The abelian bundles are defined by $c_1(V_a)= \sigma +5l-3 E_1 - 5 E_2 - E_3= -c_1(V_d), c_1(V_b)= \sigma + 5l-3 E_1 - 5 E_2 - E_3$ and $V_c$ is specified by $\eta_c=7l-3E_1-3E_2-E_3-E_4$ and $q_c=0=\zeta_c$.\footnote{Note that the fact that we are actually considering an $SU(2)$ instead of a $U(2)$ bundle is an artifact of this particular model and makes no difference in the group theoretic decomposition of $SO(32)$.}

One may verify explicitly that the conditions on $\eta_c$ for $\mu$-stability are satisfied. 
Let us also point out that the configuration is free of the of the Witten anomaly (cf.~(\ref{K-theory})). 
Furthermore, the $U(1)_Y$ hypercharge is indeed massless as desired. 
However, since the rank of the 
mass matrix is two, we get another massless $U(1)$ in the four-dimensional gauge group, 
which is identified as $U(1)_c$. 
The perturbative low energy gauge group is therefore
\bea
H=SU(3) \times SU(2) \times U(1)_Y \times U(1)'\times SO(16).
\eea

The degeneracy of the bundle $V_a$ and $V_d= V_a^*$  leads to a gauge enhancement of the $U(3)_a$ and the $U(1)_d$ to 
a $U(4)$. 
Apart from these drawbacks, the
configuration indeed gives rise to three families of the MSSM chiral
spectrum as listed in Table~\ref{MSSM}.   

\begin{table}[htb]
\renewcommand{\arraystretch}{1.5}
\begin{center}
\begin{tabular}{|c||c||c||c||c||c}
\hline
\hline
\multicolumn{5}{|c|}{$U(3)_{\a}\times U(2)_{\b} \times U(1)_{\g} \times U(1)_{\d} \times SO(16) \times \prod_a Sp(2N_a)$}\\\hline
MSSM particle & repr. & index   & mult. & total  \\\hline
$Q_L$ & $(\3,\ov{\2};1,1)_{(1,-1,0,0)}$   & $ \chi(X, V_a \otimes V_b^*)  $ & 8 &  \\
$Q_L$ & $(\3,\2;1,1)_{(1,1,0,0)}$   & $\chi(X, V_a \otimes V_b)      $  &  -11 & -3\\\hline
$u_R$ &  $(\ov{\3},1;1,1)_{(-1,0,-1,0)}$  &  $\chi(X, V_a^* \otimes V_c^*)$ & -3 & \\
$u_R$ &  $(\ov{\3},1;1,1)_{(-1,0,0,-1)}$  &  $\chi(X, V_a^* \otimes V_d^*)$ & 0  & -3 \\ \hline
$d_R$ &  $(\ov{\3},1;1,1)_{(-1,0,1,0)}$  &  $\chi(X, V_a^* \otimes V_c) $ & -3  & \\
$d_R$ &  $(\ov{\3},1;1,1)_{(-1,0,0,1)}$  &  $\chi(X, V_a^* \otimes V_d) $ & 45  &  \\  
$d_R$ & $(\ov{\3}_A,1;1,1)_{(2,0,0,0)}$  &  $  \chi(X,\bigotimes_s^2 V_a)  $ & -45 & -3\\ \hline
$L$ & $(1,\2;1,1)_{(0,1,-1,0)}$ & $ \chi(X, V_b \otimes V_c^*)$ &  -7 & \\ 
$L$ & $(1,\2;1,1)_{(0,1,0,-1)}$ & $ \chi(X, V_b \otimes V_d^*)$ &  -11 &\\  
$L$ & $(1,\ov{\2};1,1)_{(0,-1,-1,0)}$ & $ \chi(X, V_b^* \otimes V_c^*)$ & 7 &   \\ 
$L$ & $(1,\ov{\2};1,1)_{(0,-1,0,-1)}$ & $ \chi(X, V_b^* \otimes V_d^*)$ & 8 & -3 \\ \hline
$e_R$ & $(1,1;1,1)_{(0,0,2,0)}$ & $  \chi(X,\bigwedge^2 V_c)$  & 0  &     \\
$e_R$ & $(1,1;1,1)_{(0,0,0,2)}$ & $ \chi(X,\bigwedge^2 V_d) $ &  0  &\\
$e_R$ & $(1,1;1,1)_{(0,0,1,1)}$ & $\chi(X, V_c \otimes V_d) $ &  -3 & -3 \\ \hline 
$\nu_R$ & $(1,1;1,1)_{(0,0,-1,1)}$ & $\chi(X, V_c^* \otimes V_d) $ &  -3 & -3 \\
\hline \hline
\end{tabular}
\caption{\small Chiral MSSM spectrum for a four-stack model with $Q_Y= \frac{1}{6}
  Q_{\a} + \frac{1}{2}(Q_{\g} + Q_{\d})$. 
}  
\label{MSSM}
\end{center}
\end{table}
In addition, we find some chiral  exotic matter in the antisymmetric representation of 
the $U(2)$ and in the bifundamental of the $SO(16)$ with the $U(3)$ and $U(2)$, respectively.   
The chosen bundles alone do not satisfy the tadpole 
cancellation condition. However, 
the resulting tadpole class is effective, i.e. corresponds to the class of 
a veritable curve on ${\cal M}$,  and can therefore be cancelled by including H5-branes. This demonstrates the 
importance of allowing for these non-perturbative objects. We therefore find an additional symplectic gauge group 
of rank 74 including chiral bifundamental matter. 
Let us conclude this example by stating that the DUY equations can be satisfied for K\"ahler moduli inside 
the K\"ahler cone and in the perturbative regime, together with
positivity of the real part of the various gauge kinetic functions.

\begin{acknowledgement}
The contents of this talk is based on work done together with Ralph Blumenhagen and Gabriele Honecker, to whom I 
am indebted for the great pleasure of their collaboration and their support. Likewise, I am very grateful to Dieter L\"ust for many valuable discussions and support. I would like to thank the organizers of the RTN network conference
\emph{Constituents, Fundamental Forces and Symmetries of the Universe}
in Corfu, Greece for the invitation to present these ideas.
\end{acknowledgement}

\renewcommand {\bibname} {\normalsize \sc References}

{}

\end{document}